\documentclass[twocolumn,showpacs,eqsecnum,pra]{revtex4}
\usepackage{amsmath}

\usepackage{amssymb}
\usepackage{epsfig}

\begin{document}

\title{Relative entropy is an exact measure of non-Gaussianity}
\author{Paulina Marian$^{1,2}$}
\email{paulina.marian@g.unibuc.ro}
\author{ Tudor A. Marian$^{1}$}
\email{tudor.marian@g.unibuc.ro}
\affiliation{$^1$Centre for Advanced  Quantum Physics,
Department of Physics, University of Bucharest, 
R-077125 Bucharest-M\u{a}gurele, Romania\\
$^2$Department of Physical Chemistry, University of Bucharest,\\
Boulevard Regina Elisabeta 4-12, R-030018 Bucharest, Romania}


\begin{abstract}
We prove that the closest Gaussian state to an arbitrary 
$N$-mode field state through the relative entropy is built 
with the covariance matrix and the average displacement 
of the given state.  Consequently, the relative entropy 
of an $N$-mode state to its associate Gaussian one is an exact distance-type measure of non-Gaussianity. In order to illustrate this finding, we discuss the general properties of the $N$-mode Fock-diagonal states and evaluate their exact entropic amount 
of non-Gaussianity.
\end{abstract}


\maketitle

\section{Introduction}
Distance-type measures for evaluating  properties of the states involved in various protocols are at the heart of quantum information from its early days.   The distance from a given state having a specific property to a reference set of states not having it has been recognized as a measure of that property \cite{Hil,PVK,VP}. Two main problems are usually faced when defining 
a distance-type measure: choosing a convenient metric and identifying the reference set of  states. Indeed, on the one hand, the metric used should be able to distinguish properly between quantum states. Accordingly, strong candidates observing the distinguishability criterion are the relative entropy \cite{Weh,Ved} and the Bures distance \cite{Bu}, although it is not always easy to calculate them. On the other hand, a reference set 
of states as large and relevant as possible is needed in order to get accurate results. Nevertheless, a few-parameter 
characterization of the reference states is highly desirable.

Had we chosen a reliable metric and identified an appropriate 
reference set of  states, quantifying the given property could be an overwhelming extremization problem. The originally proposed properties to be quantified by distance-type measures were the nonclassicality 
of continuous-variable states \cite{Hil} and the entanglement in both discrete- and continuous-variable settings \cite{PVK,VP}. More recently, various distance-type degrees of polarization were also investigated \cite{OC}. We also mention a recent interesting proposal for a unified distance-type treatment of quantum correlations by means  of the relative entropy \cite{Modi}. Note that, in spite of not being a a true metric, the relative entropy is acceptable from the standpoint of state distinguishability 
due to the quantum Sanov theorem \cite{Ved}: The probability 
of not distinguishing the states 
${\hat \sigma}^{\prime }$ and  ${\hat \sigma}^{\prime\prime}$ 
after $\cal N$ measurements performed on $\hat \sigma^{\prime}$ is 
\begin{equation}  
P_{\cal N}({\hat \sigma}^{\prime}\to {\hat \sigma}^{\prime \prime})
=\exp{[-{\cal N} {\cal S}({\hat \sigma}^{\prime}|
{\hat \sigma}^{\prime \prime})]},
\label{sanov}
\end{equation}
where the relative entropy between the states 
${\hat \sigma}^{\prime}$ and ${\hat \sigma}^{\prime \prime}$ 
is defined as the difference
\begin{equation} 
{\cal S}(\hat \sigma^{\prime }|\hat \sigma^{\prime \prime}):={\rm Tr} [\hat \sigma^{\prime }\ln(\hat \sigma^{\prime})]-  
{\rm Tr} [\hat \sigma^{\prime }\ln(\hat \sigma^{\prime \prime})].
\label{re}
\end{equation}
The relative entropy was successfully used as a measure of entanglement for pure bipartite states providing one of the few exact and general evaluations \cite{PVK,VP}. In the mixed-state case, the tableau of exact results using relative entropy 
is much poorer.

In continuous-variable quantum information, the Gaussian states (GSs) and Gaussian trasformations play an outstanding role from both the theoretical and practical viewpoints \cite{BL,WPGCRSL}. 
On the theoretical side, the GSs have convenient phase-space representations in terms of Gaussian functions \cite{FOP}. Experimentally, they are quite accessible and were effectively obtained with light, Bose-Einstein condensates, trapped ions, 
and Josephson junctions. Although the GSs proved to be important resources in quantum information processing \cite{BL,WPGCRSL}, it was quite recently realized that non-Gaussian states 
and operations could be more efficient in some quantum protocols such as teleportation, cloning, and entanglement distillation. 
To quantify the non-Gaussianity as a resource in such cases, 
some distance-type measures of this property were proposed 
\cite{P2,P3}. However, in order to avoid the expected 
complications of the necessary extremization procedures, a more pragmatic view prevailed in defining the degrees of non-Gaussianity \cite{P2,P3,P33,Simon,GMM}. Let us designate by ${\hat \tau}_G$ 
the GS having the same mean displacement and covariance matrix 
as the given state $\hat \rho$. Instead of considering the whole reference set of the GSs, it has been convenient to choose 
${\hat \tau}_G$ as a unique reference state for all the distances employed. A measure of non-Gaussianity was then evaluated as the distance from the given state $\hat \rho$ to the associate GS 
${\hat \tau}_G$. In Refs.\cite{P2,P3,P33}, Genoni {\em et al.} introduced the Hilbert-Schmidt and relative-entropy measures 
of non-Gaussianity. Then, in Ref.\cite{Simon}, a non-Gaussianity
measure expressed as the difference between the Wehrl entropies 
of the states ${\hat \tau}_G$ and $\hat \rho$ was proposed, 
while in Refs.\cite{GMM,MGM} the non-Gaussianity based on 
the Bures metric was similarly introduced. 
 
In the present paper we choose the relative entropy to measure
the closeness of the quantum states of light. Specifically, 
the non-Gaussianity of an arbitrary $N$-mode state is reconsidered. 
Let ${\cal G}$ be the set of all the $N$-mode GSs. We define here 
the entropic amount of non-Gaussianity of a given $N$-mode state 
$\hat \rho$ as its minimal relative entropy to any $N$-mode GS:
\begin{equation}
\delta_{\cal S}[\hat \rho]:= \min_{{\hat \rho}_G \in {\cal G}}
{\cal S}(\hat \rho|{\hat \rho}_G). 
\label{qs} 
\end{equation}
The paper is organized as follows. Section II recalls the general structure of a GS. In Sec. III we prove that the closest $N$-mode GS to an arbitrary $N$-mode state is precisely the associate GS 
of the latter. Section IV is devoted to the special case of the Fock-diagonal states. In Sec. V we present our concluding remarks.

\section{Gaussian density operators}
In the following we deal with a state space that is the tensor 
product of $N$ one-mode Hilbert spaces:
\begin{equation}
{\cal H}=\bigotimes_{j=1}^{N}\,{\cal H}_j.  
\label{HS}
\end{equation} 
We start from two fundamental mathematical results: 
(i) the algebraic theorem of Williamson \cite{W,Fol} and
(ii) the existence of the metaplectic representation 
of the symplectic group ${\rm Sp}(2N, \mathbb{R})$ 
on the $N$-mode Hilbert space \eqref{HS} \cite{Met}.
As a consequence, any $N$-mode GS ${\hat \rho}_G$ can be obtained by performing a suitable unitary transformation of a well-defined 
$N$-mode thermal state (TS) $\hat \rho_T$: 
\begin{equation}
{\hat \rho}_G=\hat D(\{\alpha\})\hat U(S)
\hat {\rho}_T \hat U^{\dag}(S)\hat D^{\dag}(\{\alpha\}).
\label{G}
\end{equation} 
Recall that an $N$-mode TS is a product state
\begin{equation}
\hat \rho_T=\bigotimes_{j=1}^{N}\,(\hat \rho_T)_j({\bar n}_j)  
\label{TS}
\end{equation} 
whose one-mode factors have the exponential form
\begin{equation}
({\hat \rho}_T)_j({\bar n}_j):=\frac{1}{{\bar n}_j+1}
\exp{(-{\eta}_j {\hat a}_j^{\dagger}{\hat a}_j)},
\label{T}
\end{equation}
with ${\hat a}_j:=\frac{1}{\sqrt{2}}({\hat q}_j+i{\hat p}_j)$ 
being the annihilation operator of the mode $j$. 
In Eq.\ (\ref{T}), ${\bar n}_j$ is the Bose-Einstein 
mean photon occupancy
\begin{equation}
{\bar n}_j=[\exp{({\eta}_j)}-1]^{-1}
\label{BE}
\end{equation}
and ${\eta}_j$ is the positive dimensionless parameter 
\begin{equation}
{\eta}_j:=\frac{\hbar {\omega}_j}{k_{B}T_j}=
\ln\left(\frac{\bar n_j+1}{\bar n_j}\right).
\label{etaj}
\end{equation}
Further, in Eq.\ (\ref{G}), $\hat D(\{\alpha\}):=
\otimes_{j=1}^{N}\,{\hat D}_j(\alpha_j)$ stands for an $N$-mode Weyl displacement operator, where ${\hat D}_j({\alpha}_j):=
\exp{({\alpha}_j {\hat a}_j^{\dagger}-{\alpha}_j^* {\hat a}_j)}$. 
Then $S$ is the symplectic $2N\times 2N $ matrix, 
$S \in {\rm Sp}(2N, \mathbb{R})$, stated by the above-cited Williamson theorem, and $\hat{U}(S)$ is the corresponding unitary operator of the metaplectic representation on the $N$-mode Hilbert space \eqref{HS}. We designate the canonical quadrature operators of the modes as follows: 
\begin{equation}
\hat R_{2j-1}:=\hat q_j, \quad \hat R_{2j}:=\hat p_j
\quad (j=1, 2, \dotsc , N).
\label{R} 
\end{equation}
We employ here Einstein's summation convention and recall 
two unitary transformations of the quadratures:
\begin{align}
& \hat{U}(S)\hat R_m \hat{U}^{\dag}(S)=S_{lm}\hat R_{l},
\label{U} \\
& \hat{D}(\{\alpha\})\hat R_m \hat{D}^{\dag}(\{\alpha\})
=\hat R_m-{\bar R}_m(\{\alpha\})\hat I.
\label{D} 
\end{align}
In Eq.\ (\ref{D}), ${\bar R}_m(\{\alpha\})$ denotes the expectation
value of the quadrature $\hat R_m$ in the $N$-mode coherent state
$|\{\alpha\} \rangle \langle \{\alpha\}|$. We find it convenient 
to introduce the abbreviation
\begin{equation}
\delta \hat R_m:=\hat R_m-{\bar R}_m(\{\alpha\})\hat I
\qquad (m=1, 2, ..., 2N),
\label{dev} 
\end{equation}
as well as the $2N$-dimensional column vectors $\hat R$ 
and $\delta \hat R$, whose components are the quadrature 
operators\ (\ref{R}) and\ (\ref{dev}), respectively. 
By using the diagonal matrix
\begin{equation}
\eta:=\bigoplus_{j=1}^{N}\,{\eta}_j{\sigma}_0 
\in M_{2N}(\mathbb{R}),
\label{eta} 
\end{equation}
with ${\sigma}_0$ the $2\times 2$ identity matrix, a TS  
$\hat \rho_T$ [Eqs.\ (\ref{TS}) and\ (\ref{T})] reads
\begin{align}
\hat \rho_T &=\exp{\left[ -\sum_{j=1}^{N}\ln{(\bar n_j+1)} 
+\frac{1}{4}{\rm tr}(\eta) \right]}\notag \\ 
& \times \exp{\left( -\frac{1}{2}{\hat R}^T \eta \hat R \right)}.
\label{W} 
\end{align}
Note that we have employed the symbol ${\rm tr}$ to designate 
the trace of a matrix, while the symbol ${\rm Tr}$ denotes 
the trace of an operator on the state space\ (\ref{HS}). Substitution of Eq.\ (\ref{W}) into Eq.\ (\ref{G}) leads, 
via the transformation rules\ (\ref{U}) and\ (\ref{D}), 
to the exponential form of an arbitrary Gaussian density operator:
\begin{align}
\hat \rho_G &=\exp{\left[ -\sum_{j=1}^{N}\ln{(\bar n_j+1)} 
+\frac{1}{4}{\rm tr}(\eta) \right]}\notag \\ 
& \times \exp{\left[ -\frac{1}{2}(\delta \hat R)^T (S\eta S^T)\,
\delta \hat R \right]}.
\label{GS} 
\end{align}
Accordingly, the logarithm of any Gaussian density operator
is a quadratic polynomial in the canonical quadrature operators 
[Eq.\ (\ref{R})]:
\begin{align}
\ln{(\hat \rho_G)} &=-\frac{1}{2}(\delta \hat R)^T (S\eta S^T)\,
\delta \hat R \notag \\ 
& +\left[ -\sum_{j=1}^{N}\ln{(\bar n_j+1)} 
+\frac{1}{4}{\rm tr}(\eta) \right]\hat I. 
\label{log} 
\end{align}
Similar considerations are made in Ref. \cite{Hol}. 

It is worth stressing that there is no need to specify 
the explicit form of the unitaries $\hat{U}(S)$ in order to write
Eq.\ (\ref{log}). However, for one-mode GSs $(N=1)$,
Eq.\ (\ref{G}) reduces to their familiar parametrization as
displaced squeezed thermal states \cite{MM93}.

\section{Closest Gaussian state}

The main result of this paper is the following theorem.

{\em Theorem 1.} The nearest GS to a given $N$-mode state 
$\hat \rho$, as measured by the relative entropy, is precisely 
its associate GS ${\hat \tau}_G$. 

{\em Proof.} We exploit the two equations that define 
the associate GS ${\hat \tau}_G$ of a given $N$-mode state 
$\hat \rho$:
\begin{align}
& {\rm Tr}\left[ (\hat{\rho} -\hat{\tau}_G) {\hat R}_l \right] =0,
\label{R1} \\
& {\rm Tr}\left[ (\hat{\rho}-\hat{\tau}_G)
\frac{1}{2}\left ( {\hat R}_l{\hat R}_m
+{\hat R}_m{\hat R}_l \right ) \right ] =0.
\label{R2}
\end{align}
Owing to the quadratic stucture of the operator\ (\ref{log}), 
Eqs.\ (\ref{R1}) and\ (\ref{R2}) imply the identity
\begin{equation}
{\rm Tr}[\hat{\rho}\ln{(\hat{\rho}_G)}]
={\rm Tr}[\hat{\tau}_G\ln{(\hat{\rho}_G)}],
\label{lnrho}
\end{equation}
which holds for {\em any} $N$-mode GS ${\hat \rho}_G$. 
In particular,
\begin{equation}
{\rm Tr}[\hat{\rho}\ln{(\hat{\tau}_G)}]
={\rm Tr}[\hat{\tau}_G\ln{(\hat{\tau}_G)}]
=:-{\cal S}(\hat{\tau}_G),
\label{lntau}
\end{equation}
where ${\cal S}(\hat{\rho})$ denotes the von Neumann entropy 
of the state $\hat{\rho}$. Let us write the relative entropies 
of the given state $\hat \rho$ and its associate GS 
${\hat \tau}_G$ to an arbitrary GS ${\hat \rho}_G \in {\cal G}$:
\begin{align}
& {\cal S}(\hat{\rho}|\hat{\rho}_G)
=-{\rm Tr}[\hat{\rho}\ln{(\hat{\rho}_G)}]-{\cal S}(\hat{\rho}),
\label{Srho} \\
& {\cal S}(\hat{\tau}_G|\hat{\rho}_G)=
-{\rm Tr}[\hat{\tau}_G\ln{(\hat{\rho}_G)}]-{\cal S}(\hat{\tau}_G).
\label{StauG}
\end{align}
On account of Eqs.\ (\ref{lnrho}) and\ (\ref{lntau}),
Eq.\ (\ref{StauG}) reads
\begin{equation}
{\cal S}(\hat{\tau}_G|\hat{\rho}_G)=
-{\rm Tr}[\hat{\rho}\ln{(\hat{\rho}_G)}]
+{\rm Tr}[\hat{\rho}\ln{(\hat{\tau}_G)}].
\label{StauG1}
\end{equation}
By subtracting Eq.\ (\ref{StauG1}) from Eq.\ (\ref{Srho})
we get the formula 
\begin{align}
{\cal S}(\hat{\rho}|\hat{\rho}_G)
-{\cal S}(\hat{\tau}_G|\hat{\rho}_G) 
& =-{\rm Tr}[\hat{\rho}\ln{(\hat{\tau}_G)}]-{\cal S}(\hat{\rho})
\notag \\ & ={\cal S}(\hat{\rho}|\hat{\tau}_G).
\label{dif}
\end{align}
An equivalent form of Eq.\ (\ref{dif}) is
\begin{equation}
{\cal S}(\hat{\rho}|\hat{\rho}_G)-{\cal S}(\hat{\rho}|\hat{\tau}_G)
={\cal S}(\hat{\tau}_G|\hat{\rho}_G)\geqq 0.
\label{dif1}
\end{equation}
According to Eq.\ (\ref{dif1}), the $N$-mode GS ${\hat \tau}_G$ 
is the closest GS, ${\hat \rho}_G \in {\cal G}$, to the given 
$N$-mode state $\hat \rho$, via the relative entropy:
\begin{equation}
{\cal S}(\hat{\rho}|\hat{\tau}_G)= \min_{{\hat \rho}_G \in {\cal G}}
{\cal S}(\hat{\rho}|\hat{\rho}_G).
\label{min}
\end{equation}
This concludes the proof of Theorem 1.

{\em Corollary 1.} The entropic non-Gaussianity, defined 
by Eq.\ (\ref{qs}), coincides with the original entropic measure introduced in Ref. \cite{P3}:
\begin{equation}
\delta_{\cal S}[\hat \rho]={\cal S}(\hat{\rho}|\hat{\tau}_G).
\label{delta}
\end{equation}

In fact, we proved here that the original entropic non-Gaussianity 
has the significance of an exact entropic amount of non-Gaussianity.
If the state  $\hat \rho$ is Gaussian, then the minimal value 
\ (\ref{qs}) is reached for ${\hat \tau}_G=\hat \rho$, 
the unique state for which the relative entropy of non-Gaussianity 
$\delta_S[\hat \rho]$ [Eq.\ (\ref{qs})] vanishes.

We point out that insertion of Eq.\ (\ref{lntau}) into 
Eq.\ (\ref{dif}) allows one to express the minimal relative 
entropy
${\cal S}(\hat{\rho}|\hat{\tau}_G)$ as a difference 
of von Neumann entropies:
\begin{equation}
{\cal S}(\hat{\rho}|\hat{\tau}_G)={\cal S}(\hat{\tau}_G)
-{\cal S}(\hat{\rho}).
\label{NG}
\end{equation}
Consequently, among all the $N$-mode states with given first- and
second-order moments of the canonical quadrature variables\ (\ref{R}),
the state with maximal von Neumann entropy is the unique Gaussian one,
namely, the $N$-mode state $\hat{\tau}_G$ \cite{HSH}. 

\section{Non-Gaussianity of a Fock-diagonal state}
We designate by $\{n\}:=\{n_1,\,n_2,\dotsc ,n_N\}$ the occupancy label
of a standard $N$-mode Fock state,
\begin{equation}
|\{n\}\rangle \langle \{n\}|:=\bigotimes_{j=1}^{N}\,
{|n_j\rangle}_{j\:j}\langle n_j|.
\label{F}
\end{equation}
An $N$-mode Fock-diagonal state $\hat{\rho}_{\cal F}(\lambda)$ 
is a mixture of Fock states\ (\ref{F}):
\begin{equation}
\hat{\rho}_{\cal F}(\lambda)=\sum_{\{n\}}{\lambda}_{\{n\}}
|\{n\}\rangle \langle \{n\}|.
\label{Fd}
\end{equation}
As a matter of fact, Eq.\ (\ref{Fd}) is the spectral resolution 
of the density operator $\hat{\rho}_{\cal F}(\lambda)$, whose
eigenprojections are the Fock states\ (\ref{F}) and whose eigenvalues ${\lambda}_{\{n\}}$, besides satisfying the general requirements
$${\lambda}_{\{n\}}\geqq 0, \qquad \sum_{\{n\}}{\lambda}_{\{n\}}
=1,$$ 
are otherwise arbitrary. By the same token, Eq.\ (\ref{Fd}) 
displays the feature of $\hat{\rho}_{\cal F}(\lambda)$ of being 
a classically correlated state \cite{Luo,Modi}. 

The one-mode reductions of the Fock-diagonal state 
(FDS)\ (\ref{Fd}), are the states 
$\hat{\rho}_j(\lambda):={\rm Tr}_{\tilde{\cal H}_j}
[\hat{\rho}_{\cal F}(\lambda)]$, where the Hilbert space 
$\tilde{\cal H}_j$ is a tensor product\ (\ref{HS}) with the factor ${\cal H}_j$ omitted. They have the spectral decompositions
\begin{equation}
\hat{\rho}_j(\lambda)=\sum_{n=0}^{\infty}({\lambda}_j)_n
{|n \rangle}_{j\:j} \langle n| \quad (j=1, 2, \dotsc , N),
\label{rhoj}
\end{equation}
with the eigenvalues $$({\lambda}_j)_{n_j}
:=\sum_{\substack{\{n\}\backslash n_j}} {\lambda}_{\{n\}}
\qquad (j=1, 2, \dotsc , N).$$
In view of a general result obtained by Modi {\em et al.} 
\cite{Modi}, the nearest product state to the FDS\ (\ref{Fd}) 
is the tensor product of its one-mode marginals\ (\ref{rhoj}):
\begin{equation}
\hat{\pi}_{\rho}(\lambda):=\bigotimes_{j=1}^{N}\,\hat{\rho}_j(\lambda).
\label{pirho}
\end{equation}
The product state \eqref{pirho} is a special FDS, 
with factorized photon-number probabilities: 
\begin{equation}
\left[ \Pi{(\lambda)}\right]_{\{n\}}:
=\prod_{j=1}^{N}\,({\lambda}_j)_{n_j}. 
\label{Lambda}
\end{equation}
The relative entropy 
\begin{equation}
{\cal S}(\hat{\rho}_{\cal F}(\lambda)|\hat{\pi}_{\rho}(\lambda))
={\cal S}(\hat{\pi}_{\rho}(\lambda))
-{\cal S}(\hat{\rho}_{\cal F}(\lambda))
\label{TMI}
\end{equation}
is called the total mutual information of the state 
$\hat{\rho}_{\cal F}(\lambda)$ \cite{Modi}. 
We are ready to state the following theorem. 

{\em Theorem 2.} The nearest GS to a given $N$-mode Fock-diagonal 
state $\hat{\rho}_{\cal F}(\lambda)$, as measured by the relative entropy, is the $N$-mode TS with the same mean photon occupancies
of the modes.

{\em Proof.} It is easy to see that the mean displacements 
of any FDS are all equal to zero. For instance,
$$\langle {\hat q}_j \rangle 
=\sum_{n=0}^{\infty}({\lambda}_j)_{n\;j}
\langle n|\,{\hat q}_j\,|n {\rangle}_j=0.$$
In this manner, we evaluate both the expectation values and 
the covariances of the quadrature observables and find
\begin{align}
\langle {\hat R}_l \rangle  & ={\rm Tr}\left [ {\hat R}_l 
\hat{\rho}_{\cal F}(\lambda)\right ] =0, 
\label{Rl} \\
\sigma{({\hat R}_l,\, {\hat R}_m)} & :=
{\rm Tr}\left[ \frac{1}{2}\left ({\hat R}_l{\hat R}_m+{\hat R}_m
{\hat R}_l \right )\hat{\rho}_{\cal F}(\lambda)\right ] 
\notag \\
& -\langle {\hat R}_l \rangle
\langle {\hat R}_m \rangle =0 \qquad (l \ne m).
\label{RlRm}
\end{align}
According to Eq.\ (\ref{RlRm}), the covariance matrix (CM)
${\cal V}_{\cal F}(\lambda)$ of the FDS 
$\hat{\rho}_{\cal F}(\lambda)$ [Eq.\ (\ref{Fd})]
is the direct sum of the $2\times 2 \;\;\text{CMs}\;\; 
{\cal V}_j(\lambda)$ of the reduced single-mode states
$\hat{\rho}_j(\lambda)$ [Eq.\ (\ref{rhoj})]:
\begin{equation}
{\cal V}_{\cal F}(\lambda)=\bigoplus_{j=1}^{N}\,{\cal V}_j(\lambda).
\label{VF}
\end{equation}

We are left to find the CM ${\cal V}_j(\lambda)$ of a reduced 
one-mode state $\hat{\rho}_j(\lambda)$ [Eq.\ (\ref{rhoj})].
In order to simplify the notation, we consider a Fock-diagonal 
single-mode state
\begin{equation}
{\hat \rho}({\lambda})=\sum_{n=0}^{\infty}\,{\lambda}_n
|n \rangle \langle n|, 
\label{rho}
\end{equation}
with arbitrary photon-number probabilities ${\lambda}_n$.
This is an undisplaced state whose covariances are
\begin{align}
& \sigma{(\hat{q},\, \hat{q})}=\sum_{n=0}^{\infty}\,{\lambda}_n
\langle n|\,{\hat q}^2|n \rangle =\langle n \rangle 
+\frac{1}{2}\,,  
\label{qq} \\
& \sigma{(\hat{p},\, \hat{p})}=\sum_{n=0}^{\infty}\,{\lambda}_n
\langle n|\,{\hat p}^2|n \rangle =\langle n \rangle 
+\frac{1}{2}\,,  
\label{pp} \\
& \sigma{(\hat{q},\, \hat{p})}=\sum_{n=0}^{\infty}\,{\lambda}_n
\langle n|\,\frac{1}{2}(\hat{q}\hat{p}+\hat{p}\hat{q})\,|n \rangle =0. 
\label{pq} 
\end{align}
Accordingly, the CM ${\cal V}(\lambda)$ of the FDS
${\hat \rho}({\lambda})$ [Eq.\ (\ref{rho})] depends solely on its
mean number of photons $\langle n \rangle$ and is proportional 
to the $2\times 2$ identity matrix ${\sigma}_0:$
\begin{equation}
{\cal V}(\lambda)=\left ( \langle n \rangle +\frac{1}{2}\, \right ) {\sigma}_0, \qquad  \langle n \rangle
=\sum_{n=0}^{\infty}\,n\,{\lambda}_n.
\label{V1}
\end{equation}
Recall that a one-mode TS is Fock-diagonal and is fully determined 
by its mean photon occupancy $\bar n$ [Eq.\ (\ref{BE})] 
via the spectral representation
\begin{equation}
{\hat \rho}_T(\bar n)=\frac{1}{{\bar n}+1}\,\sum_{n=0}^{\infty}\,
\left ( \frac{\bar n}{{\bar n}+1} \right )^n |n \rangle \langle n|. 
\label{T1}
\end{equation}
Therefore, its CM 
\begin{equation}
{\cal V}_T(\bar n)=\left ( {\bar n}+\frac{1}{2}\, \right ) 
{\sigma}_0 
\label{VT1}
\end{equation}
coincides with ${\cal V}(\lambda)$ [Eq.\ (\ref{V1})] provided 
the mean numbers of photons in the two FDSs are equal: 
${\bar n}=\langle n \rangle.$ Hence the closest GS 
to the FDS\ (\ref{rho}) is the one-mode TS 
${\hat \rho}_T(\langle n \rangle)$, with the same mean photon number.

The above conclusions apply to each reduced one-mode state 
$\hat{\rho}_j(\lambda)$ [Eq.\ (\ref{rhoj})] whose CM is therefore
of the form\ (\ref{V1}): 
\begin{equation}
{\cal V}_j(\lambda)=\left ( \langle n {\rangle}_j +\frac{1}{2}\, \right ) {\sigma}_0, \quad  \langle n {\rangle}_j
=\sum_{n=0}^{\infty}\,n\,({\lambda}_j)_n.
\label{Vj}
\end{equation}
This is also the CM of a single-mode TS 
$({\hat \rho}_T)_j(\langle n {\rangle}_j)$
whose Bose-Einstein mean occupancy \eqref{BE} is
${\bar n}_j=\langle n {\rangle}_j,$
\begin{align}
({\hat \rho}_T)_j(\langle n {\rangle}_j)
& =\frac{1}{\langle n {\rangle}_j+1} \notag \\
& \times \sum_{n=0}^{\infty}\, \left ( \frac{\langle n {\rangle}_j}
{\langle n {\rangle}_j+1} \right )^n 
|n {\rangle}_{j\;j} \langle n|. 
\label{Tj}
\end{align}
Substitution of Eq.\ (\ref{Vj}) into Eq.\ (\ref{VF}) shows that
the CM of the FDS $\hat{\rho}_{\cal F}(\lambda)$ [Eq.\ (\ref{Fd})]
is diagonal:
\begin{equation}
{\cal V}_{\cal F}(\lambda)=\bigoplus_{j=1}^{N}\,
\left[ \left( \langle n {\rangle}_j +\frac{1}{2}\, \right) 
{\sigma}_0 \right].
\label{VFd}
\end{equation}
In accordance with the remark preceding Eq.\ (\ref{Tj}),
this coincides with the CM of the $N$-mode TS 
\begin{equation}
{\hat \rho}_T(\lambda):=\bigotimes_{j=1}^{N}\,
({\hat \rho}_T)_j(\langle n {\rangle}_j).
\label{TStau}
\end{equation}
The $N$-mode TS \eqref{TStau} is the associate GS of the FDS 
\eqref{Fd}. According to Theorem 1, this is the nearest GS
to the given FDS $\hat{\rho}_{\cal F}(\lambda)$, as measured 
by the relative entropy. The remark that both states have 
the same mean numbers of photons in all the $N$ modes concludes
the proof of Theorem 2.

{\em Corollary 2.} The entropic non-Gaussianity\ (\ref{delta}) 
of a FDS \eqref{Fd} can readily be evaluated via Eq.\ (\ref{NG}): 
\begin{align} 
\delta_{\cal S}[\hat{\rho}_{\cal F}(\lambda)] 
& =\sum_{j=1}^{N}\,\left[ (\langle n {\rangle}_j+1)
\ln{(\langle n {\rangle}_j+1)} 
-\langle n {\rangle}_j \ln{(\langle n {\rangle}_j)}\right] 
\notag \\
& +\sum_{\{n\}}{\lambda}_{\{n\}} \ln{\left( {\lambda}_{\{n\}}\right) }.
\label{dFDS}
\end{align}

Note that the amount of non-Gaussianity \eqref{dFDS} depends on the
photon-number distribution $\left\{{\lambda}_{\{n\}}\right\}$
of the state $\hat{\rho}_{\cal F}(\lambda)$ via its Shannon entropy 
and the corresponding mean numbers of photons in the modes.

{\em Corollary 3.} A FDS $\hat{\rho}_{\cal F}(\lambda)$ and its
nearest product state $\hat{\pi}_{\rho}(\lambda)$ have the same
closest GS through the relative entropy.

Indeed, besides having zero mean quadratures of the field modes 
\eqref{Rl}, the FDSs $\hat{\rho}_{\cal F}(\lambda)$ 
[Eq.\ (\ref{Fd})] and $\hat{\pi}_{\rho}(\lambda)$ 
[Eq.\ (\ref{pirho})] possess the same diagonal CM 
${\cal V}_{\cal F}(\lambda)$ [Eq.\ (\ref{VF})]. Accordingly, 
they have a common associate GS. In view of Theorem 2, this is 
the TS with the same mean occupancies of the modes, 
${\hat \rho}_T(\lambda)$ [Eq.\ (\ref{TStau})]. By virtue
of Theorem 1, their common associate GS \eqref{TStau} is also 
the nearest GS to both FDSs.

Moreover, the difference of the relative entropies of non-Gaussianity \eqref{NG}
\begin{equation}
{\cal S}(\hat{\rho}_{\cal F}(\lambda)|{\hat \rho}_T(\lambda))
={\cal S}({\hat \rho}_T(\lambda))
-{\cal S}(\hat{\rho}_{\cal F}(\lambda))
\label{NG1}
\end{equation}
and
\begin{equation}
{\cal S}(\hat{\pi}_{\rho}(\lambda)|{\hat \rho}_T(\lambda))
={\cal S}({\hat \rho}_T(\lambda))-{\cal S}(\hat{\pi}_{\rho}(\lambda))
\label{NG2}
\end{equation}
is equal to the total mutual information \eqref{TMI} of the FDS 
$\hat{\rho}_{\cal F}(\lambda)$: 
\begin{align}
& {\cal S}(\hat{\rho}_{\cal F}(\lambda)|{\hat \rho}_T(\lambda))
-{\cal S}(\hat{\pi}_{\rho}(\lambda)|{\hat \rho}_T(\lambda))
\notag \\
& ={\cal S}(\hat{\rho}_{\cal F}(\lambda)|\hat{\pi}_{\rho}(\lambda)).
\label{NG2a}
\end{align}
It follows that the common associate GS ${\hat \rho}_T(\lambda)$
is closer to the product state $\hat{\pi}_{\rho}(\lambda)$ than
to the given FDS $\hat{\rho}_{\cal F}(\lambda)$:
\begin{equation}
\delta_{\cal S}[\hat{\rho}_{\cal F}(\lambda)] \geqq
\delta_{\cal S}[\hat{\pi}_{\rho}(\lambda)], 
\label{NGs}
\end{equation}
with saturation of the inequality if and only if the chosen FDS 
is a product state: 
$\hat{\rho}_{\cal F}(\lambda)=\hat{\pi}_{\rho}(\lambda).$
The entropic non-Gaussianity\ (\ref{delta}) of the product state 
$\hat{\pi}_{\rho}(\lambda)$ [Eq.\ (\ref{pirho})] is a sum
of $N$ single-mode terms:
\begin{align} 
\delta_{\cal S}[\hat{\pi}_{\rho}(\lambda)] 
& =\sum_{j=1}^{N}\,\left[ (\langle n {\rangle}_j+1)
\ln{(\langle n {\rangle}_j+1)} 
-\langle n {\rangle}_j \ln{(\langle n {\rangle}_j)}\right] 
\notag \\
& +\sum_{j=1}^{N}\,\left\{ \sum_{n=0}^{\infty}\,
\left( {\lambda}_j \right)_n \ln{\left[ \left( {\lambda}_j \right)_n
\right] }\right\} .
\label{dpirho}
\end{align}

\section{Discussion and conclusions}
The importance of Theorem 1 lies in its accuracy and generality: 
It provides an {\em exact} result that holds for {\em any} state 
of the quantum radiation field. The identity between the associate GS and the nearest one, as measured by the relative entropy, gives a deep significance to the former and a quite simple recipe of evaluating the latter without any extremization hurdle. 
Concerning the proof, we mention again the central roles 
of Williamson's theorem and of the metaplectic representation, leading to the general form \eqref{G} of a GS. In addition, we  exploited just the basic properties of the relative entropy 
of being a non-negative quantity that vanishes if an only if 
the two states coincide. The fact that the entropic amount 
of non-Gaussianity is the difference of two von Neumann entropies allows one some simple evaluations and comparisons.

One may wonder about the relevance of our choice of the FDSs 
as an insightful example. However, there are several reasons 
for this preference.

(i) The set ${\cal C}_{\cal F}$ of all the $N$-mode FDSs
is the convex hull of the Fock basis in the Hilbert space 
\eqref{HS}.

(ii) Any FDS $\hat{\rho}_{\cal F}(\lambda)$ is fully determined
by its photon-number distribution 
$\left\{{\lambda}_{\{n\}}\right\} .$

(iii) The FDSs are commuting states. This property makes    calculations easier.

(iv) Any FDS has zero mean quadratures of the modes.

(v) The FDSs are classically correlated states \cite{Luo,Modi}. 
Their projective quantum discord \cite{Modi} vanishes and 
{\em a fortiori} their original quantum discords \cite{OZ}
are zero.

(vi) The nearest product state to a FDS \eqref{Fd} is the state\eqref{pirho} which also belongs to the convex set 
${\cal C}_{\cal F}.$

(vii) As shown by Theorem 2, the closest GS to a FDS \eqref{Fd}
is the TS \eqref{TStau} with the same mean numbers of photons.
This state, like any $N$-mode TS, belongs to the set 
${\cal C}_{\cal F}.$

(viii) The TSs are the only FDSs that are Gaussian.

There is one more feature of the FDSs that allows us to understand
some aspects regarding the relationship between the set of all 
the $N$-mode states and its convex subset ${\cal C}_{\cal F}$. 
Specifically, a FDS \eqref{Fd} can be prepared starting from 
any $N$-mode state $\hat \rho |_{\lambda}$ with the same  
photon-number  probabilities: 
$\langle{\{n\}}|\hat \rho |_{\lambda}|{\{n\}}\rangle 
={\lambda}_{\{n\}}.$ 
The required quantum operation is a nonselective ideal  
von Neumann-L\"{u}ders measurement \cite{L} of the numbers of photons in all the $N$ field modes. By performing it, an $N$-mode state 
$\hat \rho |_{\lambda}$ is transformed to the FDS \eqref{Fd}:
\begin{equation}
\hat{\rho}_{\cal F}(\lambda)
=\sum_{\{n\}}\,|{\{n\}}\rangle \langle{\{n\}}|\hat \rho |_{\lambda}
|{\{n\}}\rangle \langle{\{n\}}|. 
\label{qo}
\end{equation}
The nonselective projective measurement \eqref{qo} is 
a L\"{u}ders map \cite{WB} denoted by ${\Lambda}_{\{n\}}$:
\begin{equation}
\hat{\rho}_{\cal F}(\lambda)
={\Lambda}_{\{n\}}{({\hat \rho}|_{\lambda})}.
\label{LM}
\end{equation}
Note the formula
\begin{equation}
{\cal S}({\hat \rho}|_{\lambda}|\hat{\rho}_{\cal F}(\lambda)) 
={\cal S}(\hat{\rho}_{\cal F}(\lambda))
-{\cal S}({\hat \rho}|_{\lambda}). 
\label{entdif}
\end{equation}
Equation\ (\ref{entdif}) implies the following extremum property.

{\em Proposition 1.} From all the $N$-mode states with a given 
photon-number distribution, 
$\left\{{\lambda}_{\{n\}}\right\} ,$ the FDS $\hat{\rho}_{\cal F}(\lambda)$ has the maximal von Neumann entropy:
\begin{equation}
{\cal S}(\hat{\rho}_{\cal F}(\lambda))
=\max\,{\cal S}({\hat \rho}|_{\lambda}). 
\label{pnd}
\end{equation}

We specialize another general result obtained by Modi {\em et al.} 
\cite{cloF} in the form of the following statement. 

{\em Proposition 2.} Let us consider an arbitrary $N$-mode state 
$\hat \rho |_{\lambda}$ whose photon-number probabilities are ${\lambda}_{\{n\}}.$ Then the closest $N$-mode FDS to the given state $\hat \rho |_{\lambda}$ by way of the relative entropy 
is the FDS $\hat{\rho}_{\cal F}(\lambda)$ that has the same 
photon-number distribution $\left\{{\lambda}_{\{n\}}\right\},$
\begin{equation}
{\cal S}({\hat \rho}|_{\lambda}|\hat{\rho}_{\cal F}(\lambda)) 
\leqq {\cal S}({\hat \rho}|_{\lambda}|\hat{\rho}_{\cal F}(\mu)), 
\label{Fdmin}
\end{equation}
or, equivalently,
\begin{equation}
{\cal S}({\hat \rho}|_{\lambda}|\hat{\rho}_{\cal F}(\lambda)) 
=\min_{\left\{{\mu}_{\{n\}}\right\}}
{\cal S}({\hat \rho}|_{\lambda}|\hat{\rho}_{\cal F}(\mu)). 
\label{minFd}
\end{equation}
This means that the relative entropy \eqref{entdif} is 
the entropic distance from an arbitrary $N$-mode state 
${\hat \rho}|_{\lambda}$ to the convex set ${\cal C}_{\cal F}$ 
of all the $N$-mode FDSs. 

To conclude, the FDSs have some conspicuous properties and 
at the same time they can be handled rather easily. This justifies 
why we chose them as an illustrative class of field states.

\section*{ACKNOWLEDGMENT}

This work was supported  by the Romanian National Authority for Scientific Research through Grant No.~PN-II-ID-PCE-2011-3-1012 
for the University of Bucharest.


\begin{thebibliography}{50}
\bibitem{HSH} This property of the von Neumann entropy of possessing a constrained maximum was stated and proven 
in Ref. \cite{Hol}, as Proposition (i).
\end{thebibliography}
\end{document}